\documentstyle[aps,prl,floats]{revtex}
\begin{document}
\twocolumn[\hsize\textwidth\columnwidth\hsize\csname@twocolumnfalse%
\endcsname
\title{Abelian Dominance in Wilson Loops}
\author{Y.\ M.\ Cho\vskip 0.3cm}
\address{ Asia Pacific Center for Theoretical Physics,
\\and
\\Department of Physics, College of Natural Sciences, Seoul National University, Seoul 151-742
\\Korea
\\{\scriptsize \bf ymcho@yongmin.snu.ac.kr} \\ \vskip 0.3cm}
\maketitle
\date{\today}
\begin{abstract}
It has been conjectured that the Abelian projection of QCD 
is responsible for the confinement of color.
Using a gauge independent definition of the Abelian projection which
does {\it not} employ any gauge fixing, we provide a strong evidence for
the Abelian dominance in Wilson loop integral. In specific 
we prove that the gauge potential
which contributes to the Wilson loop integral is precisely the 
one restricted by the Abelian projection.  

\vspace{0.3cm}
\end{abstract}

\bigskip
]
The confinement problem in QCD is perhaps one of the most difficult problems in theoretical physics.
It has long been argued that the monopole condensation could provide the confinement of the color
through a dual Meissner effect \cite{nambu,cho1}.  More explicitly it has been conjectured
that the restricted part of QCD which comes from the ``Abelian projection" of the theory
to its maximal Abelian subgroup is responsible for the dynamics of the dual Meissner
effect\cite{cho1,cho2}.
This conjecture, which asserts that in non-Abelian gauge theory only the degrees which correspond to its maximal Abelian subgroup
should play the important role in the infra-red limit of the theory, is generally
known as the ``Abelian dominance", and has recently been readdressed by many authors
in the literature \cite{ezawa,brower}.
But the actual proof of the Abelian dominance and the monopole condensation in the low energy
limit of QCD has remained difficult.\par
A simple criterion for the confinement is given by the Wilson loop:
the ``area law'' for the vacuum expectation value of the Wilson loop for a large current
loop of a colored source, if established, could assure
the confinement.  This suggests that the Abelian dominance could be tested through
the Wilson loop calculation. In this direction, a remarkable progress has been made by the numerical simulation
during the last decade.  In fact, the lattice calculation has confirmed the Abelian
dominance, and showed that the dominant contribution to the string tension of the $q \bar q$ pair (about 92\%)
comes from the Abelian projection of the theory \cite{kronfield,stack}.  If this is so, one
should be able to prove the Abelian dominance with a field-theoretic method, independent of the numerical
simulation.  The purpose of this Letter is to provide a theoretical proof of the Abelian
dominance in the Wilson loop integral.  In specific we prove that {\it it is the 
restricted connection, more precisely the gauge invariant part of the restricted connection,
which contributes to the Wilson loop integral.}  Furthermore 
we show that {\it the vacuum expectation value of the Wilson
loop integral can be expressed as the weighted average of the loop integral of the restricted connection,
weighted by the effective action of the restricted QCD.}
This, together with the proof of the
monopole condensation in the restricted QCD, will endorse the dual
Meissner effect as the dynamical mechanism for the confinement of color in QCD.\par
Consider a non-Abelian gauge theory of a given gauge group $G$,
\begin{eqnarray}\label{eq0} {\cal L} = - \frac{1}{4} \vec{F}_{\mu \nu}^2,
\end{eqnarray}
where $\vec{F}_{\mu\nu}$ is the field strength.  To prove the Abelian dominance in the infra-red limit,
one must first clarify the meaning of the Abelian projection, and show how to project out the connection which
corresponds to the maximal Abelian subgroup $H$ from the full non-Abelian connection of
the group G.  For $SU(2)$ this means that one should tell how to separate the $U(1)$
component of the connection in a gauge-independent way.  To do this, one must select
the $U(1)$ direction at each space-time point, and make a gauge-independent projection of
the connection which contains only the $U(1)$ degree.  This can be done by introducing
a unit iso-triplet scalar field $\hat{n}(x)$ which transforms covariantly under the gauge
transformation, and insisting that $\hat{n}$ remains unchanged under the parallel
transport\cite{cho1}. So we require $\hat{n}$ to be a covariant constant,
\begin{eqnarray}\label{eq1}D_{\mu}\hat{n}=0 ~~ (\hat n ^2 =1).
\end{eqnarray}
Clearly $\hat{n}$ selects the $U(1)$ direction at each space-time point, and the parallel
transport (2) provides the desired Abelian projection of the full connection
$\vec{A}_{\mu}$,
\begin{eqnarray}\label{projecion} \vec{A}_{\mu} \longrightarrow  \hat{A}_{\mu}=
A_{\mu}\hat{n} - \frac{1}{g}\hat{n}\times\partial_{\mu}\hat{n},
\end{eqnarray}
where $ A_{\mu}=\hat{n}\cdot\vec{A}_{\mu}$ is the ``naive" Abelian component
(the electric potential) of the connection.
This $\hat{A}_{\mu}$ is the restricted connection we introduced some time ago\cite{cho1,cho2}.
It has many interesting features.  First, $\hat{n}$  being gauge covariant,
the projection (3) is obviously gauge-independent.  Moreover, $\hat{A}_\mu$ retains
the full $SU(2)$ gauge degrees of freedom even though it is clearly restricted.
This is because the gauge-independent projection still makes it an $SU(2)$ connection.
Indeed, under an arbitrary gauge transformation specified by an infinitesimal parameter $\vec{\theta}$, one has
\begin{eqnarray}{\delta}\hat{n}= -\vec{\theta}\times\hat{n},~
{\delta}\vec{A}_{\mu} = \frac{1}{g}D_\mu\vec{\theta},
\end{eqnarray}
which guarantees
\begin{eqnarray}{\delta}\hat{A}_{\mu} = \frac{1}{g}\hat{D}_\mu\vec{\theta}
= \frac{1}{g}(\partial_\mu \vec\theta + g\hat A_\mu\times\vec\theta).
\end{eqnarray}
More importantly, $\hat{A}_\mu$ retains the full topological characteristics of the original non-Abelian potential.
In fact, the isolated singularities of $\hat{n}$ defines $\pi_2(S^2)$
which describes the non-Abelian monopoles\cite{cho1,cho3}.  Indeed $\hat A_\mu$ with $A_\mu =0$ and $\hat n= \hat r$ describes precisely
the Wu-Yang monopole \cite{wu}.  Besides, with the $S^3$ compactification of $R^3$, $\hat{n}$ defines the
Hopf invariant $\pi_3(S^2)$ which describes the topologically distinct vacuua
\cite{cho4}.\par
The above discussion implies that there exists a subclass of the non-Abelian gauge theory, the restricted
gauge theory, which contains only the restricted potential which nevertheless has the full non-Abelian gauge
degrees of freedom \cite{cho1,cho2}.
To understand this, notice that with the restricted connection (3) one has
\begin{eqnarray}
& \hat{F}_{\mu\nu} = (F_{\mu\nu}+ H_{\mu\nu})\hat{n}\mbox{,}\nonumber \\
& F_{\mu\nu} = \partial_\mu A_{\nu}-\partial_{\nu}A_\mu \mbox{,}\nonumber \\
& H_{\mu\nu} = -\frac{1}{g} \hat{n}\cdot(\partial_\mu\hat{n}\times\partial_\nu\hat{n})
= \partial_\mu \tilde C_\nu-\partial_\nu \tilde C_\mu,
\end{eqnarray}
where $\tilde C_\mu$ is the monopole potential. 
To find an explicit form of $\tilde C_\mu$ let
\begin{eqnarray} S = \exp{(-t_3\gamma)}\exp{(-t_2\alpha)}\exp{(-t_3\beta) },\nonumber
\end{eqnarray}
where $t_i$ are the adjoint representation of the $SU(2)$ generators,
and let
\begin{eqnarray} \hat{n}_i & =& S^{-1} \hat{e}_i ~~~(i=1,2,3),
\end{eqnarray}
where $\hat e_1=(1,0,0),~\hat e_2=(0,1,0),~ \mbox{and}~\hat
e_3=(0,0,1)$. Now we identify $\hat{n}$ to be $\hat{n}_3$,
\begin{eqnarray}
\hat{n} =\hat{n}_3=(\sin{\alpha}\cos{\beta},~\sin{\alpha}\sin{\beta},~\cos{\alpha}).
\end{eqnarray}
Then under the gauge transformation $S$, one has
\begin{eqnarray} \hat{A}_\mu &\longrightarrow& (A_\mu + \tilde C_\mu)\hat{e}_3,\nonumber\\
\quad \hat{F}_{\mu\nu} &\longrightarrow&(F_{\mu\nu}+H_{\mu\nu})\hat{e}_3,
\end{eqnarray}
where
\begin{eqnarray}\tilde C_\mu &=& \frac{1}{g} (\cos{\alpha}\partial_\mu\beta+
\partial_\mu\gamma)\nonumber\\
&=&-\frac{1}{g}\hat{n}_1\cdot\partial_\mu\hat{n}_2.
\end{eqnarray}
This shows that $\tilde C_\mu$ describes precisely
the Dirac's monopole potential
around the isolated singularities of $\hat{n}$.  So the restricted gauge theory describes the dual dynamics of the color charge and the
non-Abelian monopole. \par
With the restricted connection it is easy to obtain the full connection $\vec{A}_{\mu}$.  Since the connection space
forms an affine space, one can obtain a most general non-Abelian connection $\vec{A}_{\mu}$
simply by adding to $\hat{A}_{\mu}$ a gauge-covariant vector field $\vec{X}_{\mu}$ which is orthogonal to $\hat{n}$,
\begin{eqnarray} \vec{A}_{\mu}&=&A_{\mu}\hat{n} - \frac{1}{g}
\hat{n}\times\partial_{\mu}\hat{n}+\vec{X}_{\mu}\nonumber\\
&=&\hat{A}_{\mu}+ \vec{X}_{\mu}. \end{eqnarray}
From (4) and (5) one can easily
confirm that $\vec{X}_\mu$ indeed forms a covariant multiplet,
\begin{eqnarray}
\delta\vec{X}_{\mu}= -\vec{\theta}\times\vec{X}_{\mu}.
\end{eqnarray}
With (11) one has
\begin{eqnarray}\label{b1}
\vec {F}_{\mu\nu} = (F_{\mu\nu}+ H_{\mu\nu}+X_{\mu\nu})\hat{n}
+\hat{D}_\mu\vec{X}_\nu- \hat{D}_\nu \vec{X}_\mu\mbox{,}
\end{eqnarray}
where $X_{\mu\nu} =g \hat{n}\cdot (\vec{X}_{\mu}\times \vec{X}_\nu)\mbox{.}$  Furthermore with
\begin{eqnarray}
\vec{X}_\mu =X^1_\mu \hat{n}_1 + X^2_\mu \hat{n}_2\mbox{,}\nonumber
\end{eqnarray}
one has
\begin{eqnarray}\label{b2}
\hat{D}_\mu \vec{X}_\nu &=&[\partial_\mu X^1_\nu-g (A_\mu+ \tilde C_\mu)X^2_\nu]\hat n_1\nonumber
\\&+&[\partial_\mu X^2_\nu+ g (A_\mu+ \tilde C_\mu)X^1_\nu]\hat{n}_2.
\end{eqnarray}
So one could express the Lagrangian explicitly in terms of $A_\mu, \hat{n}$, and $\vec{X}_\mu$,

\begin{eqnarray}{\cal L}=&-&\frac{1}{4}(F_{\mu\nu}+ H_{\mu\nu}+X_{\mu\nu})^2\nonumber
\\&-&\frac{1}{4}(\hat{D}_\mu\vec{X}_\nu-\hat{D}_\nu\vec{X}_\mu)^2.
\end{eqnarray}
Clearly this describes a $U(1)$ gauge theory coupled to a charged vector field $\vec{X}_\mu$.
But the important point here is that the $U(1)$ potential is given by $A_\mu+
\tilde C_\mu$, not just $A_\mu$.
The corresponding equations of motion is given by
\begin{eqnarray}
\partial_\mu(F_{\mu\nu}+ H_{\mu\nu}+X_{\mu\nu}) &=&-g\hat{n}\cdot \vec{X}_\mu\times(\hat{D}_\mu
\vec{X}_\nu- \hat{D}_\nu \vec X_\mu)\mbox{,}\nonumber
\\ \hat{D}_\mu(\hat{D}_\mu\vec{X}_\nu- \hat{D}_\nu \vec
X_\mu)&=&g(F_{\mu\nu}+H_{\mu\nu}\nonumber
\\&+&X_{\mu\nu})\hat{n}\times\hat{X}_\mu\mbox{.}
\end{eqnarray}
This allows us to interpret $\vec{X}_{\mu}$ as ``the valence gluon"
which plays the role of a colored source of the restricted theory \cite{cho1,cho2}.
More importantly this implies that $\vec{X}_{\mu}$, just like the quarks,
represents simply another colored source which has to be confined itself.
This is the reason why only the restricted connection (3) should
play the dominant role in the Wilson loop integral.\par
Notice that under the inverse gauge transformation $S^{-1}$ which rotates $\hat{e}_3$
to $\hat{n}$, one must have
\begin{eqnarray}
\hat{A}_\mu&=&(A_\mu+\tilde C_\mu)\hat{n}+ \frac{1}{g}{\rm tr}(-\frac{1}{2}\vec t  S^{-1}\partial_\mu
S)\nonumber
\\&=&A_\mu\hat{n}-\frac{1}{g}\hat{n}\times\partial_\mu\hat{n}\mbox{.}
\end{eqnarray}
More importantly under the gauge transformation (3) one has
\begin{eqnarray} \delta(A_\mu+\tilde C_\mu)&=& \delta(\hat{n} \cdot
\vec{A}_\mu) -\frac{1}{g} \delta(\hat{n}_1 \cdot \partial_\mu \hat
{n}_2)\nonumber\\
&=&0.
\end{eqnarray}
This shows that $(A_\mu+ \tilde C_\mu)\hat{n}$ is the gauge covariant part of the restricted
connection. This in turn allows us to identify the gauge covariant part $\vec A^{(c)}_\mu$
of the full connection,
\begin{eqnarray}
\vec{A}_\mu &=& (A_\mu+ \tilde C_\mu) \hat{n} + \vec{X}_\mu + \frac{1}{g} {\rm tr}(-\frac{1}{2} \vec{t} S^{-1} \partial_\mu S)\nonumber
\\&=&\vec A^{(c)}_\mu +  \frac{1}{g} {\rm tr}(-\frac{1}{2} \vec{t} S^{-1} \partial_\mu
S).
\end{eqnarray}
This observation will become important  in the Wilson loop calculation.\par
Now we are ready to discuss Wilson loop integral along a closed curve $\it C$.
The integral, although conceptionally simple, has remained very difficult to carry out.
But an important step to simplify the integral was made by Diakonov and Petrov\cite{diakonov},
who showed that integral for an arbitrary representation $T_i$ can
be expressed as a functional integral over all gauge transformations $S(t)$ along the loop,
\begin{eqnarray}
W(C)&=& {\rm tr}P\exp{[-\oint A^i _\mu T_i dx^\mu]}\nonumber
\\&=& \int {\cal D}S(t)\exp{[i \frac{T}{2}} \oint {\rm tr} t_3(S{\cal A}_\mu S^{-1}
\\&~&~+\frac{1}{g}S \partial_\mu S^{-1})dx^\mu],\nonumber
\end{eqnarray}
where ${\cal A}_\mu = \vec{A}_\mu \cdot \vec{t}$ and $T$ is the
Casimir invariant (i.e., the color charge) of the representation $T_i$.  Now from (19) we
have
\begin{eqnarray}
-\frac{1}{2}{\rm tr}&[&t_3 (S{\cal A}_\mu S^{-1} +\frac{1}{g}S \partial_\mu
S^{-1})]\nonumber
\\ &=&-\frac{1}{2}{\rm tr}[S^{-1}t_3 S({\cal A}_\mu-\frac{1}{g}S^{-1}\partial_\mu S)]\nonumber
\\ &=&A_\mu+\tilde C_\mu.
\end{eqnarray}
Notice that this is precisely the gauge invariant part of the restricted 
potential
\begin{eqnarray}
A_\mu+\tilde C_\mu=\hat{n}\cdot \vec A^{(c)}_\mu.
\end{eqnarray}
This makes us to identify the gauge field configuration which is relevant to the Wilson loop
integral.  {\it It is indeed the restricted gauge potential, more precisely the gauge invariant
part of the restricted potential, that contributes to the Wilson loop integral.}
This strongly endorses the Abelian dominance 
in QCD \cite{cho1,cho2}.\par
With (20) and (21) we can now obtain the desired expression for the vacuum expectation value of the Wilson loop,
\begin{eqnarray}
<W(C)>&=& \int {\cal D}A_\mu {\cal D}\hat n {\cal D}\vec X_\mu\
\exp[-\frac{1}{4}\int {\vec F _{\mu\nu}^2} d^4 x\nonumber
\\&~&~-iT\oint(A_\mu + \tilde C_\mu)dx^\mu].
\end{eqnarray}
By integrating out the $\vec X _\mu$ degrees of freedom with a proper gauge fixing, one can
express the integral as the vacuum average of (20) over the
effective Lagrangian $\hat {\cal L}_{\it eff}$ of the restricted
QCD,
\begin{eqnarray}
<W(C)>&=&\int {\cal D}A _\mu {\cal D}\hat{n} \exp[-\int \hat {\cal L}_{\it eff}  d^4 x \nonumber
\\&~&~-iT\oint(A_\mu + \tilde C_\mu)dx^\mu] \nonumber
\\&=&\int{\cal D}A _\mu {\cal D}\tilde C_\mu\exp\{-\int [\hat{\cal L}_{\it eff} \nonumber
\\&~&~+iT(A_\mu + \tilde C_\mu)j^\mu ]d^4x\},
\end{eqnarray}
where
\begin{eqnarray}
j^\mu = \int \delta^4 (x -z(t)) \frac{d z^\mu}{dt} dt, \nonumber
\end{eqnarray}
and we have changed the variable $\hat n$ by $ \tilde C_\mu$ (it is understood that ${\cal D}\tilde C_\mu$
includes the Jacobian for the change of variable). Notice that here it is the effective action,
not the bare action, of the restricted theory which appeares in the loop integral. This
effective action will contain all the dynamical informations necessary for the confinement.\par
The above result shows that {\it one can reduce the evaluation of the Wilson loop integral to the evaluation
of the generating functional of the restricted gauge theory for the gauge invariant
external current $j_\mu$}.
This is really remarkable, but perhaps not so surprising. It has been known that the evaluation
of the Wilson loop integral could be related to the evaluation of a gauge invariant part of
the generating functional which is invariant under the gauge transformation of
the external current $\vec j_\mu$\cite{diakonov}.  Our result not only confirms this
but more importantly drastically simplifies the integral, which is made possible by
identifying the precise field configuration which contributes to the Wilson loop.\par
To proceed further one must know the effective action of the restricted QCD.
The actual calculation of the effective action 
goes beyond the scope of this paper,
but fortunately one can do this at one loop level. 
With the monopole potential as the classical
background, one can show that the effective action does indeed 
provide the monopole condensation \cite{cho5}.
With this the physics behind our main result (24) becomes unmistakable.
The monopole condensation should squeeze the electric flux of the colored
source into a string,
and generate the confining potential for the colored objects.\par
The main point of the paper is to express the non-Abelian Wilson loop integral in
terms of the restricted potential.
Mathematically this amounts to reducing the non-Abelian
Stoke's integral to a quasi-Abelian Stoke's integral. There have been
many unsuccessful attempts to do this in the literature. 
We have shown how to do this with the Abelian projection. 
This by itself
does not prove the confinement. The actual proof of the confinement
should come from the calculation of the effective action of the restricted QCD \cite{cho5}.\par 
We close with the following remarks:\par
1) It must be emphasized that there is a crucial difference between
our definition of the Abelian projection and that of the others. In the
popular definition of 't Hooft, the Abelian projection is regarded as a
partial gauge fixing (called the maximal Abelian gauge) of $G/H$
degrees \cite{ezawa,brower}. In another definition which does not
employ any gauge fixing, the projection is not supposed to depend on
any particular set of field configurations \cite{faber}. In comparison
our Abelian projection (3) selects a particular set of field
configurations (the restricted potential), and is explicitly
gauge independent. Furthermore after the projection the
restricted potential still enjoys the full non-Abelian
gauge degrees of freedom. \par
2) One should keep in mind that one does {\it not} have to
have the linear potential, and the Wilson loop
integral does not have to generate the area law, to guarantee the
confinement. This is because asymtotically the  $q \bar q$ string can break 
by creating the pairs. This ``screening effect'' 
could saturate the linear potential, and thus compromize the area law.  
Only when one neglects this dynamical possibility of the pair creation
in the Wilson loop integral, one can obtain the area law.\par
3) The lattice simulations suggest that the gluon loop does not generate
a linear potential, so that the area law may not 
hold for the adjoint representations \cite{michael}.
This has been (incorrectly) attributed to the screening effect in the literature.
This appears to contradict with our result, according to 
which the quark and the gluon loops give the same integral expression
except for the color charge $T$. But this appearence is misleading, because
they contribute oppositely to the effective action. In fact one can demonstrate that,
just as in the asymtotic freedom, the gluons tend to make 
pair annihilations but the quarks tend to make
pair creations. This means that the gluons generate an ``anti-screening effect'', 
while the quarks generate the screening effect \cite{cho5}.
So our result can naturally explain why the gluon loop does not
obey the area law. But we emphasize that the correct
reason for this is not the screening effect, but the anti-screening
effect.

It must be straightforward to generalize our result to an arbitrary group \cite{cho2,cho3}.
A more detailed discussion, 
including the proof of the monopole condensation and the derivation of
the effective action for the restricted QCD,  
will be given in a forthcoming paper
\cite{cho5,cho6}.\\ \par
The author thanks Professor C. N. Yang for fruitful discussions and
encouragements, and appreciates interesting discussions with D. Diakonov, L.
Faddeev, K. Fujikawa, A. Niemi, and H. Toki.
The work is supported in part by Korean Science and Engineering
Foundation, and by the BK21 project of Ministry of Education.

\end{document}